\documentclass[twocolumn,aps,superscriptaddress,showpacs,floatfix,prc]{revtex4}
\usepackage{amssymb}
\usepackage{amsmath}
\usepackage{graphicx}
\usepackage[normalem]{ulem}
\usepackage[dvips]{color}
\usepackage{bm}
\usepackage{longtable}

\renewcommand\sout{\bgroup \color{red} \ULdepth=-.5ex \ULset}

\def\esym{$E_{\mathrm{sym}}(\rho )$~}

\def\D {$\Delta(1232)$~}
\def\rc {$\rho^{\textrm{crit}}_{\Delta^-}$~}

\begin{document}

\title{Critical Density and Impact of $\Delta (1232)$ Resonance Formation in Neutron Stars}

\author{Bao-Jun Cai}
\affiliation{Department of Physics and Astronomy, Texas A$\&$M
University-Commerce, Commerce, TX 75429-3011, USA}
\author{Farrukh J. Fattoyev}
\affiliation{Department of Physics and Astronomy, Texas A$\&$M
University-Commerce, Commerce, TX 75429-3011, USA}
\affiliation{Department of Physics and Center for Exploration of
Energy and Matter, Indiana University, Bloomington, IN, 47405, USA}
\author{Bao-An Li\footnote{%
Corresponding author: Bao-An.Li$@$tamuc.edu}}
\affiliation{Department of Physics and Astronomy, Texas A$\&$M
University-Commerce, Commerce, TX 75429-3011, USA}
\author{William G. Newton}
\affiliation{Department of Physics and Astronomy, Texas A$\&$M
University-Commerce, Commerce, TX 75429-3011, USA}
\date{\today}

\begin{abstract}
The critical densities and impact of forming \D resonances in
neutron stars are investigated within an extended nonlinear
relativistic mean-field (RMF) model. The critical densities for the
formation of four different charge states of \D are found to depend
differently on the separate kinetic and potential parts of nuclear
symmetry energy, the first example of a microphysical property of
neutron stars to do so. Moreover, they are sensitive to the
in-medium Delta mass $m_{\Delta}$ and the completely unknown
$\Delta$-$\rho$ coupling strength $g_{\rho\Delta}$. In the universal
baryon-meson coupling scheme where the respective $\Delta$-meson and
nucleon-meson coupling constants are assumed to be the same, the
critical density for the first $\Delta^-(1232)$ to appear is found
to be \rc=$(2.08\pm0.02)\rho_0$ using RMF model parameters
consistent with current constraints on all seven macroscopic
parameters usually used to characterize the equation of state (EoS)
of isospin-asymmetric nuclear matter (ANM) at saturation density
$\rho_0$. Moreover, the composition and the mass-radius relation of
neutron stars are found to depend significantly on the values of the
$g_{\rho\Delta}$ and $m_{\Delta}$.
\end{abstract}

\pacs{21.65.Ef, 21.65.Cd, 21.65.Mn, 26.60.Kp} \maketitle

\section{Introduction}
Understanding properties of \D resonances in connection with
possible pion condensation\,\cite{Mig78,Mig90,Erc88} in neutron
stars and density isomers in dense nuclear
matter\,\cite{VJ,Bog82,Liz97,Oli00,Kos97} is a longstanding
challenge of nuclear many-body physics. In fact, the role of \D
resonances in neutron stars has long been regarded as an important,
and unresolved, issue\,\cite{Sha83}. Significant works have been
carried out  to understand in-medium properties of \D resonances as
well as their effects on saturation properties of nuclear matter and
the equation of state (EoS) of dense matter using various many-body
theories and interactions: see, e.g.,
refs.\,\cite{Brown75,Oset81,Oset87, Con90,Jon92,Baldo94}. However,
compared to the numerous investigations on the possible appearance
and effects of other particles, such as hyperons and deconfined
quarks, much less effort has been devoted to the study of \D
resonances in neutron stars in recent years. This is probably
partially because of the rather high \D formation density \rc in the
core of neutron stars predicted in the seminal work by Glendenning
\textit{et al.}\,\cite{Gle85,Gle91,Gle00} using a mean-field model
with parameters well constrained by the experimental data available
at the time. Using default parameters of their model Lagrangian
leading to a symmetry energy of $E_{\mathrm{sym}}(\rho_0)=36.8$\,MeV
and its density slope $L(\rho_0)\equiv
3\rho_0{\text{d}E_{\mathrm{sym}}(\rho )}/{\text{d}\rho }|_{\rho
=\rho_0}\gtrsim 90$\,MeV at saturation density
$\rho_0$\,\cite{Gle85,Gle00,Gle91}, and using the universal
baryon-meson coupling scheme in which the nucleon-meson couplings
are set equal to the \D-meson couplings
($g_{\sigma\Delta}/g_{\sigma\textrm{N}}=g_{\omega\Delta}/g_{\omega\textrm{N}}=g_{\rho\Delta}/g_{\rho\textrm{N}}=1$),
the critical density \rc above which the first $\Delta^{-}(1232)$
appears is above $9\rho_0$. This led to the conclusion that \D
resonances played little role in the structure and composition of
neutron stars. In the same studies, the extreme importance of
symmetry energy for the formation of both hyperons and \D resonances
was emphasized. Particularly, turning off the $\Delta$-$\rho$
coupling which contributes the potential part of the symmetry energy
(retaining thus kinetic symmetry energy only), the \rc is about
$3\rho_0$\,\cite{Gle85}.

Interest has been renewed with recent studies using different
symmetry energies and/or assumptions about the baryon-meson coupling
constants which have found that the \rc can be as low as $\rho_0$
and the inclusion of the \D has significant effects on both the
composition and structure of neutron
stars\,\cite{Xia03,ChY07,Sch10,Lav10,Dra14,Dra14a}. These studies
generally use some individual sets of model parameters leading to
macroscopic properties of ANM at saturation density consistent with
most if not all of the existing experimental constraints.

During the last three decades, much progress has been made in
constraining the EoS of dense neutron-rich nuclear matter. In
particular, reasonably tight constraints on the density dependence
of the nuclear symmetry energy \esym especially around the
saturation density have been obtained in recent years, see, e.g.,
refs.\,\cite{EPJA,lkb98,liudo,Bar05,LCK,Trau12,Tsang12,LiBA13,Hor14,Jim13}
for comprehensive  reviews. For example, the 2013 global averages of
the magnitude and slope of the \esym at $\rho_0$ are respectively
$E_{\mathrm{sym}}(\rho_0)=31.6\pm 2.7$\,MeV and $L=58.9\pm
16.5$\,MeV based on 28 analyses of various terrestrial laboratory
experiments and astrophysical observations\,\cite{LiBA13}. Moreover,
\D resonances play a very important role in heavy ion collisions,
see e.g., ref.\,\cite{HIon} for reviews, especially for the
production of particles such as pions, kaons and various exotic
heavy mesons. In particular, the masses of \D resonances primarily
created in nucleon-nucleon (NN) collisions through the
$\textrm{NN}\rightarrow \textrm{N}\Delta$ process act as an energy
reservoir for sub-threshold particle production. The release of this
energy in subsequent collisions involving \D resonances may help
create new particles that can not be produced otherwise in the
direct, first-chance  NN collisions. Thus, particle production has
been widely used in probing in-medium properties of \D resonances.
Since \D resonances and nucleons have an isospin 3/2 and 1/2
respectively, the total isospin is 1 or 2 for the N$\Delta$ while it
is 1 or 0 for the NN state. Because of the isospin conservation, the
\D production can only happen in the total isospin 1 NN channel.
Therefore, the abundances and properties of \D resonances are
sensitive to the isospin asymmetry of the system as neutron-neutron
pairs always have an isospin 1 while neutron-proton pairs can have
an isospin 1 or 0. 

Naturally, both heavy-ion collisions and neutron
stars are places where the isovector properties and interactions of
\D resonances are expected to play a significant role. Indeed,
useful information about the symmetry energy of dense neutron-rich
matter has been extracted from studying pion and kaon productions in
heavy-ion collisions\,\cite{Bar05,LCK}. It is especially worth
noting that the isovector (symmetry) potential of \D resonances was
recently found to affect appreciably the ratio of charged pions in
transport model simulations of heavy-ion collisions at intermediate
energies\,\cite{LiLisa}. However, to our best knowledge, no
quantitative information about the isovector interaction of \D
resonances has been extracted yet from any terrestrial experiments. On the other hand, there are strong
indications from both theoretical calculations and phenomenological
model analyses of electron-nucleus, photoabsorption and pion-nucleus
scattering that the \D isoscalar potential $V_{\Delta}$ (real part
of its isoscalar self-energy $\Sigma_{\mathrm{S}}$) is in the range
of $-30\,\mathrm{MeV}+V_{\textrm{N}}\le V_{\Delta}\le
V_{\textrm{N}}$ with respect to the nucleon isoscalar potential
$V_{\textrm{N}}$\,\cite{Dra14a}. The in-medium masses and widths of \D resonances are also the focuses of 
many experimental and theoretical studies using various reactions and techniques. To our best knowledge, however,
no clear consensus has been reached yet. For instance, from analyzing the $(p,\pi)$ invariant masses in the final state of heavy-ion collisions at SIS/GSI 
energies, indications were found for an approximately $-60$ MeV mass shift for \D resonances at the freeze-out of about 1/3 the saturation density \cite{Esk98}. 
However, photo absorption data and some advanced model calculations found no evidence of significant in-medium \D mass shift \cite{Kor04,Hee05}.
It is thus exciting that new proposals to experimentally study at
FAIR/GSI the \D resonance spectroscopy and interactions in
neutron-rich matter are being considered by the NUSTAR
Collaboration\,\cite{Lenske}.

Similar to the appearance of any other new hadron above its
production threshold in neutron stars,  the addition of \D
resonances will soften the EoS and influence the composition of
neutron
stars\,\cite{Gle85,Gle00,Gle91,Xia03,ChY07,Sch10,Dra14,Dra14a,Lav10}.
Because of charge neutrality, depending on the individual
populations of the four different charge states of \D resonances,
the density dependence of the proton fraction in neutron stars may
be modified. Then different cooling mechanisms sensitive to the
proton fraction may come into play above certain densities.
Moreover, the formation of \D resonances may also push up critical
densities for the appearance of various hyperons\,\cite{Dra14}. As
noticed earlier in the literature and emphasized in
ref.\,\cite{Sha83}, there are many interesting questions regarding
properties of \D resonances in dense matter and their impact on
observables of neutron stars. Obviously, answers to all of these
questions naturally rely on the critical density of \D formation in
dense neutron star matter.

In this work,  we first identify analytically key microphysics
quantities determining the critical formation densities of the four
charge states of \D resonances. Then, within a nonlinear RMF model
we calculate consistently the \rc as a function of the \D mass
$m_{\Delta}$, the isovector $\Delta$-$\rho$ coupling strength
$g_{\rho\Delta}$ and seven macroscopic variables characterizing the
EoS of ANM at $\rho_0$ all within their latest constraints. Finally,
impacts of the \D formation on the composition and mass-radius
correlation of neutron stars are studied.

\section{Key microphysics determining the Delta formation density in neutron stars}
\renewcommand*\figurename{\small Figure}
\begin{figure}[tbh]
\centering
  \includegraphics[width=7.5cm]{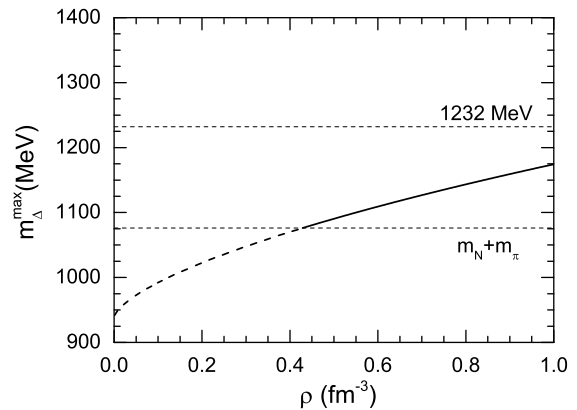}
\caption{The maximum mass of Delta resonances produced in the $\textrm{N}\textrm{N}\rightarrow\textrm{N}\Delta$ process in a free Fermi gas of nucleons at
density $\rho$.} \label{MaxDeltaMass}
\end{figure}

To set a reference for our following studies we first estimate the
\rc in a free Fermi gas of nucleons. For the head-on collision of
two nucleons both with Fermi momentum
$|\textbf{k}|=k_{\textrm{F}}=(3\pi^2\rho/2)^{1/3}$ in the
$\textrm{N}\textrm{N}\rightarrow\textrm{N}\Delta$ process, the
maximum mass of the produced \D resonance is
$m_{\Delta}^{\max}=2({k_{\textrm{F}}^2+m_{\textrm{N}}^{2}})^{1/2}-m_{\textrm{N}}
$ where $m_{\textrm{N}}$ is the average nucleon mass in free-space.
It is well known that \D has a Breit-Wigner mass distribution around
the centroid $m^0_{\Delta}=1232$\,MeV with a width of about 120 MeV.
The distribution starts at a minimum of $m_{\Delta}^{\min}\equiv
m_{\textrm{N}}+m_{\pi}\simeq 1076$\,MeV  where $m_{\pi}$ is the pion
mass. Shown in Fig. \ref{MaxDeltaMass} is the $m_{\Delta}^{\max}$
reachable in the $\textrm{N}\textrm{N}\rightarrow\textrm{N}\Delta$
process as a function of density. From this simple estimate, where
effects of the nuclear potentials are neglected, the critical
density $\rho_{\Delta}^{\textrm{crit}}$ for producing the lightest
\D resonance is about $3\rho_0$. To reach the centroid
$m^0_{\Delta}=1232$\,MeV, the density has to be far above
1\,$\mathrm{fm}^{-3}$. This estimate also illustrates the importance
of considering the mass dependence of the \D formation density in
more realistic calculations for matter in neutron stars.

In interacting nuclear systems, the masses of \D resonances and
their critical formation densities depend on the in-medium
self-energies of all particles involved. Assuming neutron stars are
made of neutrons, protons, \D resonances, electrons and muons, i.e.,
the npe$\mu\Delta$ matter in chemical and $\beta$ equilibrium, the
total baryon number density is
\begin{equation}
\rho=\rho_{\textrm{p}}+\rho_{\textrm{n}}+\rho_{\Delta^{++}}+\rho_{\Delta^{+}}
+\rho_{\Delta^{0}}+\rho_{\Delta^{-}}
\end{equation}
where
\begin{align}
\rho_{\textrm{n}}=&\frac{\left(k_{\textrm{F}}^{\textrm{n}}\right)^3}{3\pi^2},~~\rho_{\textrm{p}}=\frac{\left(k_{\textrm{F}}^{\textrm{p}}\right)^3}{3\pi^2},\\
\rho_{\Delta^{++}}=&\frac{(k_{\textrm{F}}^{\Delta^{++}})^3}{\pi^2},~~\rho_{\Delta^{-}}=\frac{(k_{\textrm{F}}^{\Delta^{-}})^3}{\pi^2},\\
\rho_{\Delta^{+}}=&\frac{(k_{\textrm{F}}^{\Delta^{+}})^3}{3\pi^2},~~\rho_{\Delta^{0}}=\frac{(k_{\textrm{F}}^{\Delta^{0}})^3}{3\pi^2}
\end{align}
with $k_{\textrm{F}}^j$'s
($j=\textrm{p,n},\Delta^{++},\Delta^{+},\Delta^{0},\Delta^{-}$)
being the corresponding Fermi momenta. The chemical equilibrium
condition for reactions
$\textrm{n}\to\textrm{p}+\textrm{e}+\overline{\nu}_{\textrm{e}}$ and
$ \textrm{p}+\textrm{e}\to\textrm{n}+\nu_{\textrm{e}}$ requires
$
\mu_{\textrm{e}}=\mu_{\textrm{n}}-\mu_{\textrm{p}}
$
where
$\mu_{\textrm{e}}=[m_{\textrm{e}}^2+(k_{\textrm{F}}^{\textrm{e}})^2]^{1/2}=[m_{\textrm{e}}^2+(3\pi^2\rho
x_{\textrm{e}})^{2/3}]^{1/2}\simeq (3\pi^2\rho
x_{\textrm{e}})^{1/3}$ with $x_{\textrm{e}}\equiv
\rho_{\textrm{e}}/\rho$ the electron fraction. When the chemical
potential of electron is larger than the static mass of a muon,
reactions $ \textrm{e}\to\mu+\nu_{\textrm{e}}+\overline{\nu}_{\mu}$,
$ \textrm{p}+\mu\to\textrm{n}+\nu_{\mu}$ and
$\textrm{n}\to\textrm{p}+\mu+\overline{\nu}_{\mu}$ will also take
place. The latter requires
\begin{equation}\label{ceq2} \mu_{\textrm{n}}-\mu_{\textrm{p}}=\mu_{\mu}=\sqrt{m_{\mu}^2+(3\pi^2\rho
x_{\mu})^{2/3}}
\end{equation} besides
$\mu_{\textrm{n}}-\mu_{\textrm{p}}=\mu_{\textrm{e}}$,
where $m_{\mu}=105.7\,\textrm{MeV}$ is the mass of a muon and
$x_{\mu}\equiv \rho_{\mu}/\rho$ is the muon fraction. On the other
hand, the following four types of inelastic reactions will take place between
nucleons and the four charge states of Delta resonances
\begin{align}
\Delta^{++}+\textrm{n}\longleftrightarrow\textrm{p}+\textrm{p},\\
\Delta^++\textrm{n}\longleftrightarrow\textrm{n}+\textrm{p},\\
\Delta^0+\textrm{p}\longleftrightarrow\textrm{p}+\textrm{n},\\
\Delta^-+\textrm{p}\longleftrightarrow\textrm{n}+\textrm{n}.
\end{align}
Their chemical equilibrium requires then
\begin{align}
\mu_{\Delta^{++}}=&2\mu_{\textrm{p}}-\mu_{\textrm{n}},\label{ceq3}\\
\mu_{\Delta^{+}}=&\mu_{\textrm{p}},\label{ceq4}\\
\mu_{\Delta^{0}}=&\mu_{\textrm{n}},\label{ceq5}\\
\mu_{\Delta^{-}}=&2\mu_{\textrm{n}}-\mu_{\textrm{p}}.\label{ceq6}
\end{align}
In addition, the total charge neutrality in neutron stars requires that
$
x_{\textrm{p}}+x_{\Delta^+}+2x_{\Delta^{++}}=x_{\textrm{e}}+x_{\mu}+x_{\Delta^-}
$
where $x_{\Delta^-}\equiv \rho_{\Delta^-}/\rho$, $x_{\Delta^+}\equiv
\rho_{\Delta^+}/\rho$, and $x_{\Delta^{++}}\equiv
\rho_{\Delta^{++}}/\rho$, respectively.

Within the framework of a given nuclear many-body theory, the Eqs.
(\ref{ceq3})-(\ref{ceq6}) can be used to calculate the critical
formation densities for the four charge states of \D resonances.
Generally speaking, in relativistic mean-field models, a baryon of
bare mass $m_{\mathrm{baryon}}$ obtains a Dirac effective mass
$m_{\textrm{dirac}}^{\ast}(\mathrm{baryon})=m_{\mathrm{baryon}}+\Sigma_{\textrm{S}}$
and a chemical potential
$\mu_{\mathrm{baryon}}=[{k_{\textrm{F}}^2+m_{\textrm{dirac}}^{\ast
2}}(\mathrm{baryon})]^{1/2}+\Sigma_{\textrm{V}}$ where
$\Sigma_{\textrm{S}}$ and $\Sigma_{\textrm{V}}$ are the real parts
of its scalar and vector self-energies, respectively. Consider the
$\Delta^-$ formation, for example, noticing that
$\mu_{\textrm{n}}-\mu_{\textrm{p}}\simeq4E_{\textrm{sym}}(\rho)\delta$
and using non-relativistic kinematics, the Eq. (\ref{ceq6}) leads to
the following condition for producing a $\Delta^-$ of bare mass
$m_{\Delta^-}$ at rest
\begin{align}\label{Dcondition}
&\frac{(k_{\textrm{F}}^{\textrm{n}})^2}{2m_{\textrm{dirac}}^{\ast}}=\frac{\left[3{\pi}^2(1+\delta)\rho/2\right]^{2/3}}{2m_{\textrm{dirac}}^{\ast}}\\
&=m_{\Delta^-}-m_{\textrm{N}}-4E_{\textrm{sym}}(\rho)\delta+\Sigma^{\Delta}_{\textrm{S}}-\Sigma^{\textrm{N}}_{\textrm{S}}+\Sigma^{\Delta^{-}}_{\textrm{V}}
-\Sigma^{\textrm{n}}_{\textrm{V}}\notag
\end{align}
where $m^{\ast}_{\textrm{dirac}}$ is the nucleon Dirac effective
mass and $\delta=(\rho_{\textrm{n}}-\rho_{\textrm{p}})/\rho$ is the
isospin asymmetry of nucleons before \D resonances are produced.
Given the density dependencies of the symmetry energy and
self-energies, this equation determines the \rc in neutron stars at
$\beta$ equilibrium. It also shows clearly what microphysics
quantities determine the \rc. In particular, the difference in Delta
and nucleon masses $m_{\Delta^-}-m_{\textrm{N}}$, the symmetry
energy \esym, the difference in both scalar
$\Sigma^{\Delta}_{\textrm{S}}-\Sigma^{\textrm{N}}_{\textrm{S}}$ and
vector
$\Sigma^{\Delta^{-}}_{\textrm{V}}-\Sigma^{\textrm{n}}_{\textrm{V}}$
self-energies are all characteristics of baryon interactions. It
also indicates where the model dependence and uncertainties are. As
we mentioned earlier, experimental data indicates that the
difference in nucleon and \D isoscalar self-energies can be up to
about 30\,MeV while there is simply no experimental indication so
far about the difference in their isovector self-energies. We notice
that in many studies in the literature the \D resonances and
nucleons are assumed to have the same scalar and vector
self-energies. In this case then, the \rc is completely determined
by the \D mass $m_{\Delta}$ and the nuclear symmetry energy as a
function of density \esym.

The nonlinear RMF model has been very successful in describing many
nuclear properties and phenomena during the last few decades, see
e.g.,
refs.\,\cite{Ser86,Rei89,Rin96,Men06,Sug94,Lal97,Lon04,Jia10,Fat10,Fat10a,Agr12,Fat13,Mul96,Hor01,Tod05,Che07,Cai12,Cai14,Cai14a}.
The total Lagrangian density of the nonlinear RMF model of
ref.\,\cite{Hor01} augmented by the Yukawa couplings of the Delta
fields to various isoscalar and isovector meson fields, can be
written
as\,\cite{Gle85,Bog82,Kos97,Liz97,Oli00,Xia03,Lav10,Dra14,Dra14a}
\begin{align}\label{NLRMF}
\mathcal{L}=&\overline{\psi}_{\textrm{N}}\big[\gamma_{\mu}(i\partial^{\mu}-g_{\omega\textrm{N}}\omega^{\mu}
-g_{\rho\textrm{N}}\vec{\mkern1mu\tau}_{\textrm{N}}\cdot\vec{\mkern1mu\rho}^{\mu})\notag\\
&\hspace*{4cm}-(m_{\textrm{N}}-g_{\sigma\textrm{N}}\sigma)\big]\psi_{\textrm{N}}\notag\\
&+\overline{\psi}_{\Delta\nu}\big[\gamma_{\mu}(i\partial^{\mu}-g_{\omega\Delta}\omega^{\mu}
-g_{\rho\Delta}\vec{\mkern1mu\tau}_{\Delta}\cdot\vec{\mkern1mu\rho}^{\mu})\notag\\
&\hspace*{4cm}-(m_{\Delta}-g_{\sigma\Delta}\sigma)\big]\psi_{\Delta}^{\nu}\notag\\
&+\frac{1}{2}\partial_{\mu}\sigma\partial^{\mu}\sigma-\frac{1}{2}m_{\sigma}^2\sigma^2-U_{\textrm{N}}(\sigma)\notag\\
&+\frac{1}{2}m_{\omega}^2\omega_{\mu}\omega^{\mu}
-\frac{1}{4}\omega_{\mu\nu}\omega^{\mu\nu}
+\frac{1}{4}c_{\omega\textrm{N}}(g_{\omega\textrm{N}}^2\omega_{\mu}\omega^{\mu})^2\notag\\
&+\frac{1}{2}m_{\rho}^2\vec{\mkern1mu\rho}_{\mu}\cdot\vec{\mkern1mu\rho}^{\mu}-\frac{1}{4}\vec{\mkern1mu\rho}_{\mu\nu}\cdot\vec{\mkern1mu\rho}^{\mu\nu}\notag\\
&+
\frac{1}{2}\left(g_{\rho\textrm{N}}^2\vec{\mkern1mu\rho}_{\mu}\cdot\vec{\mkern1mu\rho}^{\mu}\right)\Lambda_{\textrm{V}}g_{\omega\textrm{N}}^2\omega_{\mu}\omega^{\mu},
\end{align}
where $\omega_{\mu \nu }\equiv \partial _{\mu }\omega _{\nu
}-\partial _{\nu
}\omega _{\mu }$ and $\rho_{\mu \nu }\equiv \partial _{\mu }\vec{%
\mkern1mu\rho }_{\nu }-\partial _{\nu }\vec{\mkern1mu\rho }_{\mu }$
are strength tensors for $\omega$ and $\rho$ meson fields,
respectively. $\psi_{\textrm{N}}$, $\psi_{\Delta}^{\nu}$, $\sigma $,
$\omega _{\mu }$, $\vec{\mkern1mu\rho }_{\mu }$ are the nucleon
Dirac field, Schwinger-Rarita field for $\Delta$ resonances,
isoscalar-scalar meson field, isoscalar-vector meson field and
isovector-vector meson field, respectively, and the arrows denote
isovectors,
$U_{\textrm{N}}({\sigma})=b_{\sigma\textrm{N}}m_{\textrm{N}}(g_{\sigma\textrm{N}}\sigma)^3/3+c_{\sigma\textrm{N}}(g_{\sigma\textrm{N}}\sigma)^4/4
$ is the self interaction term of the $\sigma$ field. The parameter
$\Lambda _{\textrm{V}}$ represents the coupling constant of mixed
interaction between the isovector $\rho$ and isoscalar $\omega$
mesons. It is known to be important for calculating the density
dependence of the symmetry energy\,\cite{Hor01}. In terms of the
expectation values of the meson fields, $\overline{\sigma}$,
$\overline{\omega}_0$ and $\overline{\rho}_0^{(3)}$ where the
subscript ``0" denotes the zeroth component of the four-vector while
the superscript  ``(3)" denotes the third component of isospin, the
nucleon and \D isoscalar self-energies are respectively
\begin{align}
\Sigma^{\textrm{N}}_{\textrm{S}}&=-g_{\sigma\textrm{N}}\overline{\sigma}\\
\mathrm{and}~
\Sigma^{\Delta}_{\textrm{S}}&=-g_{\sigma\Delta}\overline{\sigma}.
\end{align}
Their isovector self-energies are respectively
\begin{align}
\Sigma^{\textrm{N}}_{\textrm{V}}&=g_{\omega\textrm{N}}\overline{\omega}_0
+\tau^3_{\textrm{p/n}}g_{\rho\textrm{N}}\overline{\rho}_0^{(3)}\\
\mathrm{and}~ \Sigma^{\Delta}_{\textrm{V}}&=g_{\omega\Delta}\overline{\omega}_0
+\tau^3_{i}g_{\rho\Delta}\overline{\rho}_0^{(3)}
\end{align}
with $\tau^3_{\textrm{p}}=+1$, $\tau^3_{\textrm{n}}=-1$ and
$i=\Delta^{++}$, $\Delta^+$, $\Delta^0$, $\Delta^-$,
$\tau^3_{\Delta^{++}}=+3$, $\tau^3_{\Delta^{+}}=+1$,
$\tau^3_{\Delta^{0}}=-1$, $\tau^3_{\Delta^{-}}=-3$.

In terms of the ratios of Delta-meson over nucleon-meson coupling
constants $x_{\sigma}\equiv g_{\sigma\Delta}/g_{\sigma\textrm{N}}$,
$x_{\omega}\equiv g_{\omega\Delta}/g_{\omega\textrm{N}}$ and
$x_{\rho}\equiv g_{\rho\Delta}/g_{\rho\textrm{N}}$, the
Eqs.(\ref{ceq3})-(\ref{ceq6}) lead to the following conditions
determining the critical densities for forming the four charge
states of \D resonances
\begin{widetext}
\begin{align}
\rho_{\Delta^-}^{\textrm{crit}}:&~~\frac{\left(k^{\textrm{n}}_{\textrm{F}}\right)^2}{2m_{\textrm{dirac}}^{\ast}}\simeq
\Phi_{\Delta}+g_{\sigma \textrm{N}}(1-x_{\sigma})\overline{\sigma}
-g_{\omega\textrm{N}}(1-x_{\omega})\overline{\omega}_0-6(1-x_{\rho})E_{\mathrm{sym}}^{\textrm{pot}}(\rho)\delta
-4E_{\textrm{sym}}^{\textrm{kin}}(\rho)\delta,\label{def1}\\
\rho_{\Delta^0}^{\textrm{crit}}:&~~\frac{\left(k^{\textrm{n}}_{\textrm{F}}\right)^2}{2m_{\textrm{dirac}}^{\ast}}\simeq
\Phi_{\Delta}+g_{\sigma \textrm{N}}(1-x_{\sigma})\overline{\sigma}
-g_{\omega \textrm{N}}(1-x_{\omega})\overline{\omega}_0-2(1-x_{\rho})E_{\mathrm{sym}}^{\textrm{pot}}(\rho)\delta,\label{def2}\\
\rho_{\Delta^+}^{\textrm{crit}}:&~~\frac{\left(k^{\textrm{p}}_{\textrm{F}}\right)^2}{2m_{\textrm{dirac}}^{\ast}}\simeq
\Phi_{\Delta}+g_{\sigma \textrm{N}}(1-x_{\sigma})\overline{\sigma}
-g_{\omega \textrm{N}}(1-x_{\omega})\overline{\omega}_0+2(1-x_{\rho})E_{\mathrm{sym}}^{\textrm{pot}}(\rho)\delta,\label{def3}\\
\rho_{\Delta^{++}}^{\textrm{crit}}:&~~\frac{\left(k^{\textrm{p}}_{\textrm{F}}\right)^2}{2m_{\textrm{dirac}}^{\ast}}\simeq
\Phi_{\Delta}+g_{\sigma \textrm{N}}(1-x_{\sigma})\overline{\sigma}
-g_{\omega\textrm{N}}(1-x_{\omega})\overline{\omega}_0+6(1-x_{\rho})E_{\mathrm{sym}}^{\textrm{pot}}(\rho)\delta
+4E_{\textrm{sym}}^{\textrm{kin}}(\rho)\label{def4}
\end{align}
where $ \Phi_{\Delta}\equiv m_{\Delta}-m_{\textrm{N}}$ is the
Delta-nucleon mass difference,
$E_{\textrm{sym}}^{\textrm{pot}}(\rho) =2^{-1}\rho
g_{\rho\textrm{N}}^2(m_{\rho}^2+\Lambda_{\textrm{V}}g_{\rho\textrm{N}}^2
g_{\omega\textrm{N}}^2\overline{\omega}_0^2)^{-1}$ and
$E_{\textrm{sym}}^{\textrm{kin}}(\rho)=6^{-1}k_{\textrm{F}}^2(k_{\textrm{F}}^2+m_{\textrm{dirac}}^{\ast,2})^{-1/2}$
are respectively the potential and kinetic part of the symmetry
energy in the nonlinear RMF model\,\cite{LCK}.
\end{widetext}
Several interesting conclusions can be made qualitatively from
inspecting the above four conditions. Generally, the critical
densities depend differently on the three coupling ratios
$x_{\sigma}$, $x_{\omega}$ and $x_{\rho}$ as they have different
natures. The isoscalar coupling ratios $x_{\sigma}$ and $x_{\omega}$
affect the four \D resonances the same way, i.e., the $x_{\sigma}$
lowers while the $x_{\omega}$ raises their critical formation
densities, while the isovector coupling ratio $x_{\rho}$ acts
differently on the four different charge states of \D resonances.
Moreover, the kinetic and potential parts of the symmetry energy
have separate and different effects. In particular, in the universal
baryon-meson coupling scheme, i.e.,
$x_{\sigma}=x_{\omega}=x_{\rho}=1$, the critical densities for
creating $\Delta^-$ and $\Delta^{++}$ depend only on the $
\Phi_{\Delta}$ and the kinetic symmetry energy
$E_{\textrm{sym}}^{\textrm{kin}}(\rho)$ besides the
$m^{\ast}_{\textrm{dirac}}$, while those for the $\Delta^0$ and
$\Delta^+$ are determined only by the $ \Phi_{\Delta}$ and
$m^{\ast}_{\textrm{dirac}}$. In this case, assuming the
$E_{\textrm{sym}}^{\textrm{kin}}(\rho)$ is always positive as in the
case of RMF, noticing that the Fermi surface of protons is lower
than that of neutrons at any density in neutron-rich matter, i.e.,
$k_{\textrm{F}}^{\textrm{p}}<k_{\textrm{F}}^{\textrm{n}}$, one then
sees immediately the following sequence of appearance
$\rho_{\Delta^-}^{\textrm{crit}}<\rho_{\Delta^0}^{\textrm{crit}}<\rho_{\Delta^+}^{\textrm{crit}}<\rho_{\Delta^{++}}^{\textrm{crit}}$\,\cite{Gle85,Dra14a}.
However, we notice that the short-range nucleon-nucleon correlation
(SRC)\,\cite{Fra81,Arr12,Hen14} may lead to negative kinetic
symmetry energies even at normal density of nuclear matter, see e.g.
refs.\,\cite{XuLi,Vid11,Lov11,Car12,Rio14,OrHen14,LiBA14}. In this
case, the order of appearance of $\Delta^-$ and $\Delta^{++}$, thus
the fraction of various particles and the structure of neutron stars
may be different. We remark here that this is the first time that
some physics quantities in neutron stars are found to depend
separately on the kinetic and potential parts instead of the total
symmetry energy.
\renewcommand*\figurename{\small Figure}
\begin{figure}[tbh]
\centering
  \includegraphics[width=8.5cm]{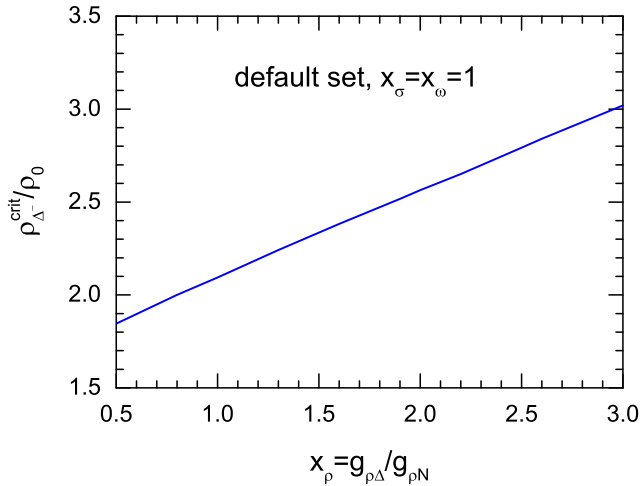}
\caption{(Color Online) Dependence of the critical density \rc for
$\Delta^-$ formation in neutron stars on the relative $\Delta-\rho$
coupling strength $x_{\rho}=g_{\rho\Delta}/g_{\rho \textrm{N}}$.}
\label{xrho}
\end{figure}

Some earlier studies, see, e.g.,
refs.\,\cite{Liz97,Kos97,Dra14,Dra14a}, indicate that
$x_{\sigma}\approx x_{\omega}\approx 1.0$. Effects of slight
deviations from this value on properties of neutron stars have also
been reported, see, e.g., ref.\,\cite{Sch10}. To our best knowledge,
however, little is known about the range of $x_{\rho}$ and its
effects in either heavy-ion collisions\,\cite{LiLisa} or neutron
stars. Moreover, most studies so far are limited to density isomers
due to \D formation in symmetric nuclear matter where it is
sufficient to consider only effects of the $x_{\sigma}$ and
$x_{\omega}$.

In the universal baryon-meson coupling scheme, the default set
(SH-NJ) of RMF model parameters leads to the following values of
macroscopic quantities characterizing the EoS of ANM at
$\rho_0=0.149\,\textrm{fm}^{-3}$: the binding energy
$E_0(\rho_0)=-16.09\,\textrm{MeV}$, the Dirac effective mass
$m_{\textrm{dirac}}^{\ast0}(\rho_0)/m_{\textrm{N}}=0.64$, the
incompressibility $K_0(\rho_0)=230\,\textrm{MeV}$, the skewness
coefficient $J_0(\rho_0)=-415\,\textrm{MeV}$, the magnitude
$E_{\textrm{sym}}(\rho_0)=31.17\,\textrm{MeV}$ and slope
$L(\rho_0)=48.64\,\textrm{MeV}$ of symmetry energy. We remark that
almost all of these bulk parameters are extracted from a fit to the
properties of finite nuclei\,\cite{Cai14a} with the exception of the
skewness coefficient $J_0$ that has been obtained by fixing the
maximum mass of a neutron star at $M_{\rm max} = 2.01
M_{\odot}$\,\cite{Antoniadis2013}. Variations around this
parameterization will be investigated in the following.

For Delta resonances of mass $m_{\Delta}=1232$\,MeV, we study in
Fig.\ \ref{xrho} the \rc dependence on the $x_{\rho}$  while keeping
all other quantities at their default values. It is seen that the
\rc increases approximately linearly with $x_{\rho}$. At
$x_{\rho}=1$, the \rc is only about $2.1\rho_0$ consistent with that
found in refs.\,\cite{Sch10,Dra14,Dra14a} but much smaller than the
one found by Glendenning \textit{et al.}\,\cite{Gle85,Gle00,Gle91}.
However, unless the value of $x_{\rho}$ is somehow constrained, the
\rc will remain underdetermined.

\renewcommand*\figurename{\small Figure}
\begin{figure}[tbh]
\centering
  \includegraphics[width=8.5cm]{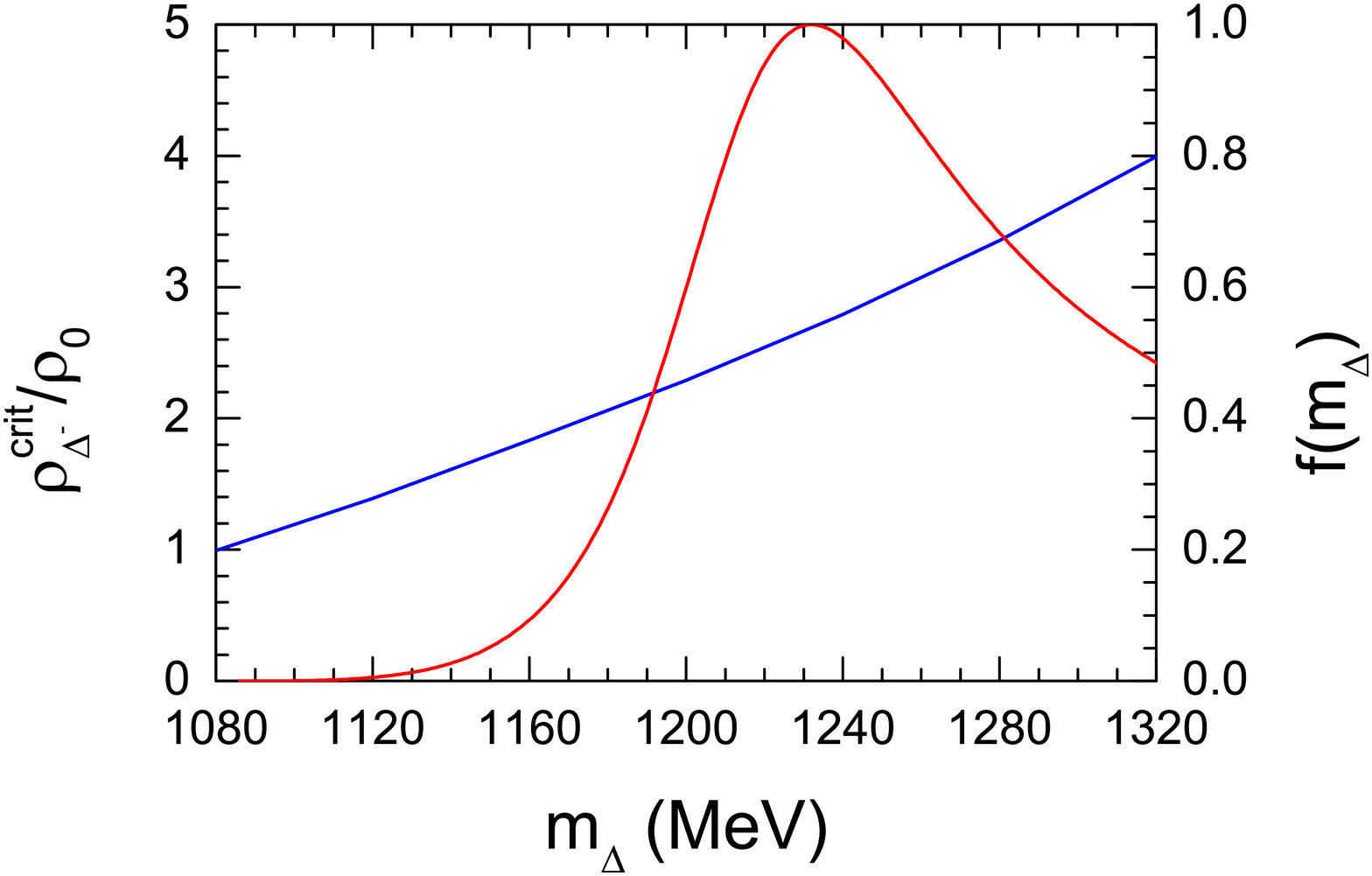}
\caption{(Color Online) The Delta mass $m_{\Delta}$ dependence of
the critical density \rc for $\Delta^-$ formation in neutron stars
(blue) and the Breit-Wigner mass distribution of Delta resonances in
free-space (red).} \label{Dmass}
\end{figure}
Delta resonances in free-space have the Breit-Wigner mass
distribution
\begin{equation}\label{fmass}
f(m_{\Delta})=\frac{1}{4}\frac{\Gamma^2(m_{\Delta})}{(m_{\Delta}-m^0_{\Delta})^2+\Gamma^2(m_{\Delta})/4}
\end{equation}
with the mass-dependent width given by\,\cite{Kit,Li93}
\begin{equation}\label{wDelta}
\Gamma(m_{\Delta}) =0.47 q^{3}/ \left(m_{\pi}^{2}+0.6q^{2}\right)~{\rm (GeV)}
\end{equation}
where
$q=[([{m^2_{\Delta}-m^2_{\textrm{N}}+m^2_{\pi}}]/{2m_{\Delta}})^2-m^2_{\pi}]^{1/2}$
is the pion momentum in the $\Delta$ rest frame in the
$\Delta\rightarrow \pi+\textrm{N}$ decay process. Shown in red in
Fig. \ref{Dmass} is the free-space \D mass distribution. It is known
that the \D mass distributions may be modified in nuclear
medium\,\cite{Lenske}. This effect is beyond the scope of the RMF
model used here as it does not consider the imaginary part of the \D
self-energy self-consistently. However, we can examine how the \rc
depends on the bare \D mass $m_{\Delta}$ by varying its value in the
Eqs. (\ref{def1})-(\ref{def4}). As one expects and indicated by the
Eqs. (\ref{def1})-(\ref{def4}), the \rc increases with $m_{\Delta}$
in the universal coupling scheme. Our numerical calculations shown
with the blue line indicate that the increase is almost linear.
Considering the mass distribution, while the \rc for $\Delta$
resonances around $m^0_{\Delta}$ is about $2.1\rho_0$, it gradually
decreases for lower $\Delta$ masses. Of course, these low-mass
$\Delta$ resonances are less likely to be produced compared to the
ones near $m^0_{\Delta}$. On the other hand, the $\Delta$ mean
lifetime $\tau_{\Delta}=\hbar /\Gamma(m_{\Delta})$ is only about 1.7
fm/c at $m^0_{\Delta}$ but increases very quickly for lower masses.
Thus, the main population of $\Delta$ resonances in neutron stars
may not necessarily peak at $m_{\Delta}=1232$\,MeV. A detailed study
of this issue will require a full account of the
$\pi-\textrm{N}-\Delta$ dynamics in neutron stars that is also
beyond the scope of the RMF model used here. Nevertheless, our
results indicate that the appearance of $\Delta$ resonances,
especially the ones with low masses around $2\rho_0$, may compete
with other particles, such as hyperons, thus possibly modify the
widely accepted and longtime viewpoint that hyperons should appear
earlier than \D resonances in neutron stars\,\cite{Gle85,Hae07}. To
our best knowledge, however, no study to date has considered
consistently effects of the mass distribution and the associated
mass-dependent lifetimes of $\Delta$ resonances in neutron stars.

\begin{figure*}[tbh]
\begin{center}
\includegraphics[width=6.5in]{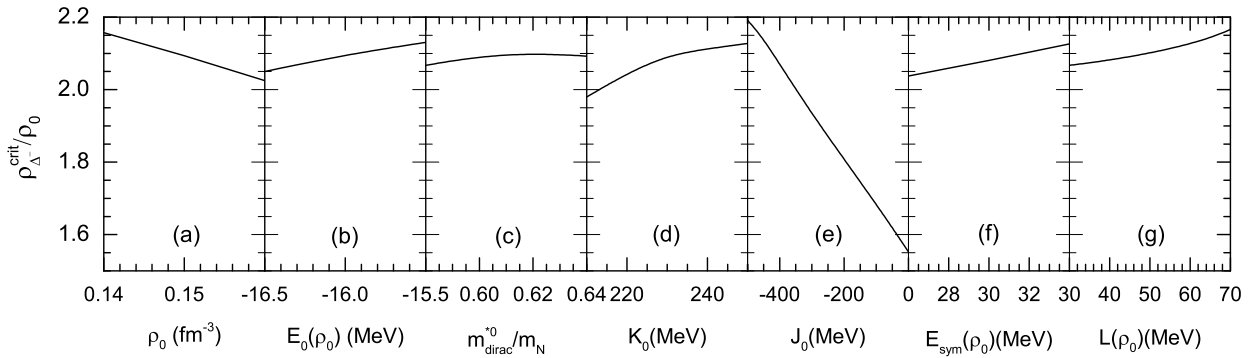}
\caption{Critical density for the formation of $\Delta^{-}$
resonance from the nonlinear RMF model by varying individually
$\rho_0$ (a), $E_0(\rho_0)$ (b), $m_{\mathrm{dirac}}^{\ast 0}$ (c),
$K_{0}$ (d), $J_{0}$ (e), $E_{\textrm{sym}}(\rho_0)$ (f) and
$L(\rho_0)$ (g).} \label{CDen}
\end{center}
\end{figure*}

\section{Effects of nuclear equation of state on the formation of \D resonances in neutron stars}
The equation (\ref{ceq6}) for determining the critical density
$\rho^{\textrm{crit}}_{\Delta^-}$ can be rewritten as
\begin{align}\label{minc}
\mu_{\Delta^{-}}^{\min}=&m_{\Delta}+\Sigma^{\Delta^-}
=2\mu_{\textrm{n}}-\mu_{\textrm{p}}\simeq\mu_{\textrm{n}}+4E_{\textrm{sym}}(\rho)\delta.
\end{align}
The left hand side is given in terms of the microscopic quantities,
i.e., \D mass and the three Delta-meson coupling constants in the
total self-energy
$\Sigma^{\Delta^-}=-g_{\sigma\Delta}\overline{\sigma}+g_{\omega\Delta}\overline{\omega}_0
-3g_{\rho\Delta}\overline{\rho}_0^{(3)}$. Using the parabolic
approximation for the EoS of ANM $E(\rho,\delta)\simeq
E_0(\rho)+E_{\textrm{sym}}(\rho)\delta^2+\mathcal{O}(\delta^4)$ and
assuming the density is not too far from $\rho_0$, the right hand
side of Eq. (\ref{minc}) can be expanded in terms of $\chi =(\rho
-\rho _{0})/3\rho _{0}$ and the isospin asymmetry $\delta$ as
\begin{align}\label{eq46}
\mu_{\Delta^-}^{\min}\simeq& m_{\textrm{N}}+E_0(\rho_0)
+\left(\frac{\chi}{3}\frac{\rho}{\rho_0}+\frac{\chi^2}{2}\right)K_0\notag\\
&+\frac{\chi^2}{6}\left(\frac{\rho}{\rho_0}+\chi\right)J_0
+(2\delta+3\delta^2)E_{\textrm{sym}}(\rho_0)\notag\\
&+\left[\frac{\rho}{3\rho_0}\delta^2+\chi(2\delta+3\delta^2)\right]L(\rho_0)
\end{align}
where higher order terms in the expansion have been neglected (for
full details see ref.\,\cite{Cai14}). Here $K_{0} =\left. 9\rho
_{0}^{2}{\text{d} ^{2}E_{0}(\rho )}/{\text{d} \rho ^{2}}\right\vert
_{\rho =\rho _{0}}$ and $J_{0} =\left. 27\rho _{0}^{3}{\text{d}
^{3}E_{0}(\rho )}/{\text{d} \rho ^{3}}\right\vert _{\rho =\rho
_{0}}$ are the incompressibility and skewness coefficient of
symmetric nuclear matter (SNM) at $\rho_0$, respectively. This
expansion is very easy to understand considering the energy
conservation in the Delta production process $\textrm{NN}\rightarrow
\textrm{N}\Delta$, i.e., the minimum energy of the Delta is the
energy of a nucleon (sum of nucleon rest mass and its mechanical
energy). Since all seven macroscopic quantities used to characterize
the EoS of ANM i.e., (a) the saturation density $\rho_0$ of SNM
where the pressure $P(\rho _{0})=0$, (b) the binding energy
$E_{0}(\rho_{0})$, (c) incompressibility $K_{0}$, (d) skewness
coefficient $J_{0}$, (e) nucleon effective mass
$m_{\mathrm{dirac}}^{\ast 0}$, (f) magnitude $
E_{\textrm{sym}}(\rho_0)$ and (g) slope $L(\rho_0)$ of the symmetry
energy, are all explicit functions (see, e.g.,
refs.\,\cite{Cai14,Cai12a,Che14a}) of the seven RMF microscopic
model parameters, i.e., $g_{\sigma\textrm{N}}$,
$g_{\omega\textrm{N}}$, $g_{\rho\textrm{N}}$,
$b_{\sigma\textrm{N}}$, $c_{\sigma\textrm{N}}$,
$c_{\omega\textrm{N}}$ and $\Lambda_{\textrm{V}}$, the Eq.
(\ref{eq46}) allows us to explore the \rc dependence on either the
seven microscopic or macroscopic parameters. Since the macroscopic
quantities are either empirical properties of nuclear matter or
directly related to experimental observables, it is more useful for
the purposes of this work to examine the \rc by varying individually
the seven macroscopic quantities. We notice that within both the RMF
and Skyrme-Hartree-Fock approaches similar correlation
analyses\,\cite{Che10}  have been successfully applied to study the
neutron skin\,\cite{Che10,Zha13}, the giant monopole resonances
(GMR) of finite nuclei\,\cite{Che12}, the higher-order bulk
characteristic parameters of ANM\,\cite{Che11a}, the electric dipole
polarizability $\alpha_{\textrm{D}}$ in
$^{208}\textrm{Pb}$\,\cite{Zha14}, the correlation between the
maximum mass of neutron stars and the skewness coefficient of the
SNM\,\cite{Cai14}, as well as the relationship between the \esym and
the symmetry energy coefficient in the mass formula of
nuclei\,\cite{Che11}.

\begin{figure}[tbh]
\centering
  \includegraphics[width=8.cm]{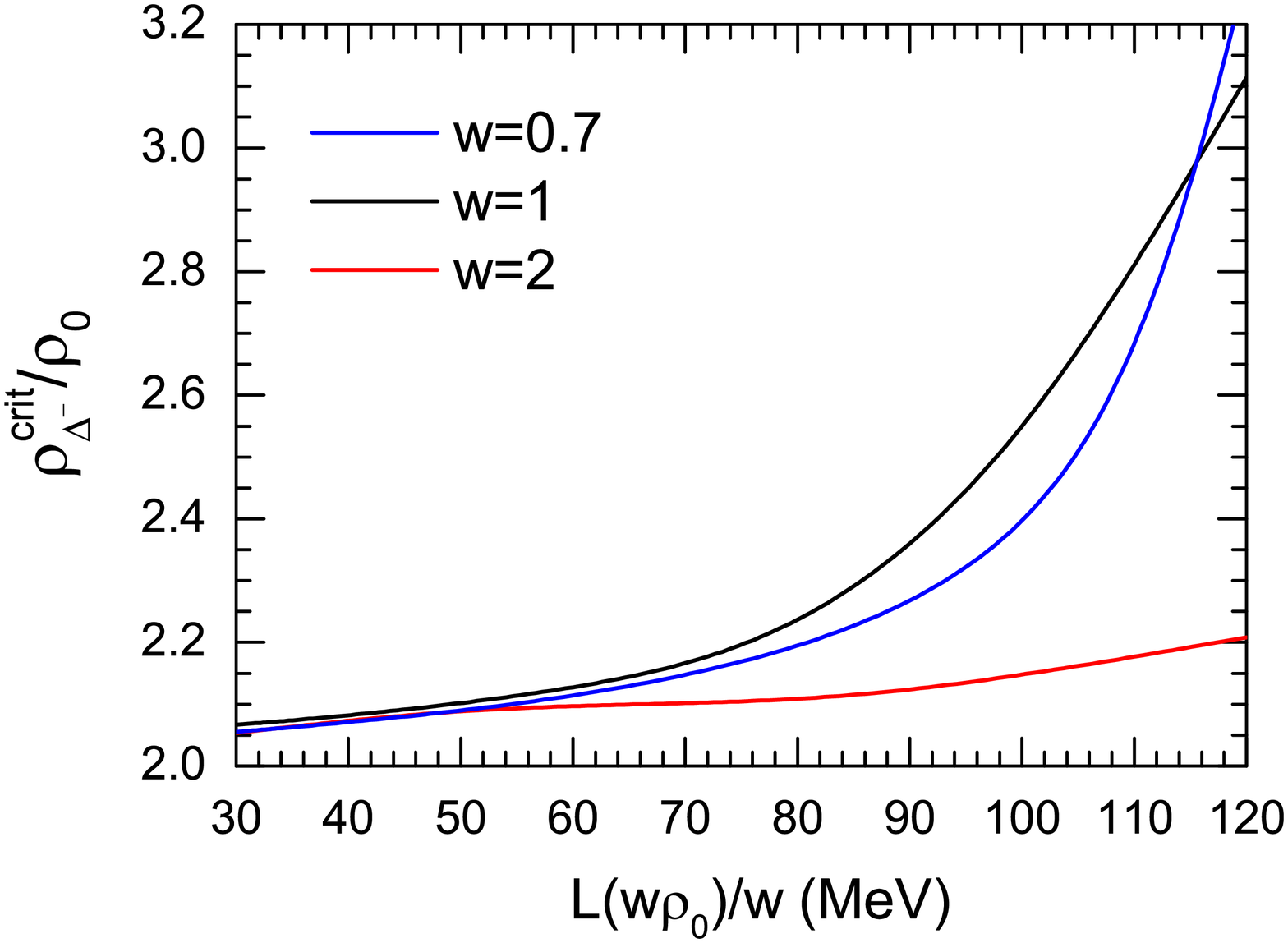}
\caption{(Color Online) Critical density for the formation of
$\Delta^{-}$ resonance as a function of the scaled density slope of
symmetry energy $L(w\rho_0)/w$ at three reduced densities of
$\rho_{\textrm{r}}/\rho_0=w=0.7, 1$ and 2, respectively}
\label{CDenL}
\end{figure}

In Fig.\ \ref{CDen} we show the \rc obtained in the universal
coupling scheme as a function of the seven macroscopic parameters
within their respective uncertain ranges. Among them,  the
$J_0=-250\pm 250\,\textrm{MeV}$ and $L=50\pm20\,\textrm{MeV}$
currently have the largest uncertainties. Examining the results
shown in Fig. \ref{CDenL}, we make the following two observations:
(i) The values of \rc from all seven correlations overlap around
\rc=$(2.08\pm0.02)\rho_0$. Notice that this result depends on the
value of $J_0$ in the default set, which had been fixed to obtain
the maximum stellar mass configuration matching the observed $2.01
M_{\odot}$ neutron star. With this assumption then this result is
the most reliable prediction for the \rc consistent with all of the
constraints within the RMF model considered. We note that increasing
$J_0$ raises the maximum mass and lowers the critical density \rc;
hence the range \rc=$(2.08\pm0.02)\rho_0$ constitutes an upper limit
on \rc. Since the inclusion of Delta resonances inevitably softens
the EOS, the magnitude of $J_0$ should therefore be much larger than
the one used in our default set to be consistent with the current
observation of the two solar-mass neutron star. With the other six
macroscopic quantities fixed within their current uncertainty ranges, then the
critical density \rc will take an even smaller value than the upper
limit predicted above. And this is one of the main reasons why we
haven't considered hyperons in the present work, because they appear
at much higher densities than \rc. (ii) A reasonable and
quantitative measure of the sensitivity of \rc to each individual
variable is the response function $\mathcal{Q}\equiv
\left|(\textrm{d}y/y)/(\textrm{d}x/x)\right|$ where $\textrm{d}y/y$
is the relative change in \rc with respect to its mean value and
$\textrm{d}x/x$ is the relative change in the variable x with
respect to its mean value. The value of $\mathcal{Q}$ is
approximately $\mathcal{Q}[E_{0}(\rho_{0})]=0.53$,
$\mathcal{Q}[\rho_0]=0.40$, $\mathcal{Q}[K_{0}]=0.29$,
$\mathcal{Q}[m_{\mathrm{dirac}}^{\ast 0}]=0.18$,
$\mathcal{Q}[J_{0}]=0.17$,
$\mathcal{Q}[E_{\textrm{sym}}(\rho_0)]=0.11$ and
$\mathcal{Q}[L(\rho_0)]=0.10$ following the same order of importance
in the expansion of the minimum Delta chemical potential in Eq.
(\ref{eq46}).

\renewcommand*\figurename{\small Figure}
The seemingly stronger correlation between the
$\rho_{\Delta^-}^{\textrm{crit}}$ and $J_0$ compared to the
correlations of \rc with the other 6 variables is because of the
relatively larger uncertainty in $J_0$. 
On the other hand, some of the weaker correlations shown in Fig.\
\ref{CDen} may become much stronger if one goes beyond their current
constraints. For example, shown in Fig. \ref{CDenL} are the
correlations of the \rc with the reduced slope
$L(\rho_{\textrm{r}})/w$ of \esym at three reference densities of
$\rho_{\textrm{r}}/\rho_0=w=0.7, 1.0$ and 2 while fixing the
magnitudes of \esym and other variables at their default values. It
is seen that the \rc increases much faster for
$L(\rho_{\textrm{r}})/w\ge 70$\,MeV. This feature is consistent with
that found in ref.\,\cite{Dra14a}. Noting again that some earlier
studies have used models predicting $L$ values high than 90\,MeV,
they thus predicted correspondingly much large values for the \rc.

\section{Effects of Delta resonances on the composition and structure of neutron stars}
Having shown that the \rc depends sensitively on the completely
unknown $\Delta$-$\rho$ coupling strength $g_{\rho\Delta}$ and the
Delta mass $m_{\Delta}$, and it is about \rc=$(2.08\pm0.02)\rho_0$
in the universal coupling scheme for $m_{\Delta}=1232$\,MeV using
RMF model parameters consistent with all existing constraints on the
nuclear EoS, we now turn to effects of Delta formation on properties
of neutron stars. This study is carried out within the npe$\mu\Delta$ model omitting other
particles such as hyperons and quarks at high densities. This model is
sufficient for the purposes of this work. Moreover, we restrict ourselves
to studying effects of the Delta formation on the composition and
mass-radius relation of neutron stars by varying the Delta mass and
its coupling strength with the $\rho$ meson. In constructing the EoS
of various layers in neutron stars for solving the
Oppenheimer-Volkoff (OV) equation, we follow a rather standard
scheme.  For the core  we use the EoS of $\beta$-stable and charge
neutral npe$\mu\Delta$ matter obtained from the nonlinear RMF model
described earlier. The inner crust with densities ranging between
$\rho_{\text{out}}=2.46\times 10^{-4}$ fm$^{-3}$ corresponding to
the neutron dripline and the core-crust transition density $\rho
_{\text{t}}$ determined self-consistently using the thermodynamical
method\,\cite{Cai12,XuJ09} is the region where complex and exotic
nuclear structure ---collectively referred to as the ``nuclear
pasta" may exist.  Because of our poor knowledge about this region
we adopt the polytropic EoSs parameterized in terms of the pressure
$P$ as a function of total energy density $\varepsilon$ according to
$P=a+b\varepsilon^{4/3}$\,\cite{XuJ09,Hor03}. The constants $a$ and
$b$ are determined by the pressure and energy density at $\rho
_{\text{t}}$ and $\rho _{\text{out}}$\,\cite{XuJ09}. For the outer
crust \cite{BPS71}, we use the BPS EoS for the region with
$6.93\times 10^{-13}$\,fm$^{-3}<\rho <\rho _{\text{out}}$ and the
FMT EoS for $4.73\times 10^{-15}$\,fm$^{-3}<\rho <$$6.93\times
10^{-13}$\,fm$^{-3}$, respectively.

\renewcommand*\figurename{\small Figure}
\begin{figure}[tbh]
\centering
  \includegraphics[width=8.cm]{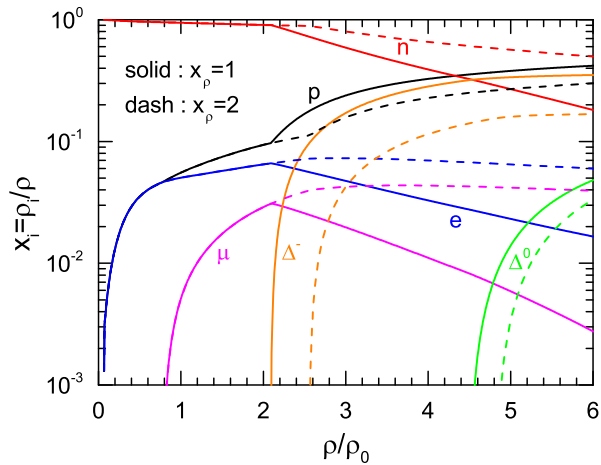}
\caption{(Color Online) Fractions of different species in neutron
stars composed of neutrons, protons, Delta resonances, electrons and
muons within the nonlinear RMF model using two sets of model
parameters with $x_{\rho}=g_{\rho\Delta}/g_{\rho \textrm{N}}=1$ and
2, respectively.} \label{Yi}
\end{figure}

Shown in Fig. \ref{Yi} are fractions of different species, i.e.,
$x_i=\rho_i/\rho$, in neutron stars using two parameter sets with
the $\Delta$-$\rho$ coupling strength corresponding to $x_{\rho}=1$
and 2, respectively. These calculations are done with
$m_{\Delta}=1232$\,MeV. It is seen that the appearance of $\Delta^-$
affects significantly the fractions of others particles depending on
the value of the $x_{\rho}$ as one expects. The modified fractions
of the lighter particles $e$ and $\mu$ will affect their weak decays
and thus the possible kaon condensation. Moreover, the strong boost
of the proton fraction above the $\Delta^-$ production threshold may
have a significant impact on the cooling processes in neutron
stars\,\cite{Yak01}. A detailed investigation of these consequences
requires a self-consistent extension of the model considered here
and is on the agenda of our future work.

\renewcommand*\figurename{\small Figure}
\begin{figure}[tbh]
\centering
  \includegraphics[width=8.cm]{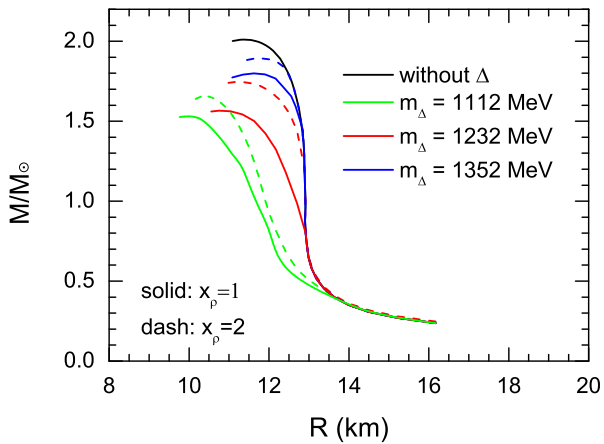}
\caption{(Color Online) The mass-radius correlation of neutron stars
without and with Delta resonances of different masses using
$x_{\rho}=g_{\rho\Delta}/g_{\rho \textrm{N}}=1$ and 2,
respectively.} \label{MR1}
\end{figure}

Finally, we study in Fig.\ \ref{MR1} effects of both the Delta mass
$m_{\Delta}$ and the $x_{\rho}$ parameter on the mass-radius
relation of neutron stars.  Without including the Delta resonances (black line), as we mentioned earlier the default value of
$J_0$ was chosen to predict a maximum neutron star mass of $2.01M_{\odot}$ consistent with the latest observations.
As  given in Eq. (\ref{fmass}) and Eq. (\ref{wDelta}), \D has an intrinsic mass distribution and each mass has its own width (lifetime) in free space. 
Especially, it is possible to form \D with masses much smaller than the centroid of 1232 MeV in low energy $NN\rightarrow N\Delta$ or $\pi+N\rightarrow \Delta$
reactions. As one expects, including Delta resonances reduces both the maximum mass and the corresponding
radius. Moreover, the effect is stronger for Delta resonances with
masses smaller than the centroid. These low mass Delta resonances have lower production thresholds and are thus more abundant. They then
softens the EoS more starting at lower densities reached in the cores of lighter neutron stars. Of course, effects of the Delta resonates on properties of neutron stars depend on the 
value of $x_{\rho}$. As shown in Fig.\ \ref{Yi} and discussed earlier, a larger value of $x_{\rho}$ leads
to a higher \rc. Thus, the effect of varying $m_{\Delta}$ appears at
a higher mass (higher central density) of neutron stars when the
value of $x_{\rho}$ is increased from 1 to 2.

Obviously, including the Delta resonances we can no longer obtain a maximum neutron star mass of $2.01M_{\odot}$ with the default parameter set. Moreover, we found that by varying the value of $J_0$ in its uncertain range, we can not recover a maximum mass of $2.01M_{\odot}$ with $x_{\rho}=1$. We confirm that indeed there is a ``Delta puzzle" consistent with the finding of refs.\ \cite{Dra14,Dra14a}. Thus, if indeed Delta resonances can be formed around $2\rho_0$ as in the universal coupling scheme, our findings call for serious considerations of the various EoS stiffening mechanisms proposed in the literature. On the other hand, as we mentioned earlier, there is currently no constraint on the $x_{\rho}$ neither theoretically nor experimentally.
Since the critical density for Delta formation increases approximately linearly with $x_{\rho}$, it is therefore critical to first constrain independently the $x_{\rho}$, such as from heavy-ion collisions, before the ``Delta puzzle" in understanding properties of neutron stars can be resolved.

\section{Summary and Discussions}
The possible formation and roles of \D resonances in neutron stars
are outstanding issues in nuclear astrophysics. The first and most important piece of
information necessary for resolving these issues is the critical
density above which the \D can be formed in neutron stars. Previous
studies have indicated that the critical densities range from
$\rho_0$ to very high values only reachable in the core of very
massive neutron stars. In this work, within the extended nonlinear
RMF model we found that the critical formation densities for the
four different charge states of \D resonances depend differently on
the separate kinetic and potential parts of the nuclear symmetry
energy, the in-medium Delta mass $m_{\Delta}$ and the completely
unknown $\Delta$-$\rho$ coupling strength $g_{\rho\Delta}$.
\emph{This is the first time a microphysical property of neutron
star matter has been shown to depend differently on the potential
and kinetic parts of the symmetry energy.} Assuming that the
respective $\Delta$-meson and nucleon-meson coupling constants are
the same, the critical density for the first $\Delta^-(1232)$ to
appear is found to be \rc=$(2.08\pm0.02)\rho_0$ using RMF model
parameters consistent with current constraints on all seven
macroscopic parameters characterizing the EoS of ANM at $\rho_0$.
We also found that the composition and the mass-radius relation of neutron
stars are significantly affected by the formation of \D resonances.
In particular, the impacts of the \D formation depend sensitively on
the values of the $g_{\rho\Delta}$ and the in-medium Delta mass
$m_{\Delta}$ which are also being probed with terrestrial laboratory
experiments.

To this end, it is interesting to note that since the early work by Kubis and Kutschera \cite{Kub97},
the isovector, Lorentz-scalar $\delta$ meson has been incorporated in several RMF models in predicting the EOS of dense neutron-rich matter and its applications 
in understanding properties of relativistic heavy-ion collisions and neutron stars \cite{Liu02,Gre03,Men04,Gai04,Ala10,Ala10a,Pai09,Niu10,Hua15}.
Since the $\delta$ contributes negatively to the symmetry energy in contrast to the $\rho$ meson, 
to fit the saturation properties of nuclear matter, reproduce the known symmetry energy of about 31 MeV at saturation density and avoid the bounding of pure neutron matter, the $\rho$-nucleon coupling has to be readjusted significantly compared to its value in the RMF model 
without considering the $\delta$ meson since there is no known mechanism to determine the $\delta$-nucleon coupling independently. The net effect of including the $\delta$ meson is thus to stiffen the symmetry energy at supra-saturation densities where the $\rho$ contribution dominates. Indeed, applications of these models have found that the $\delta$ may play an appreciable role in determining the 
hadron-quark phase transition \cite{Ala10,Ala10a},  the crust-core transition and the formation of a pasta phase \cite{Pai09}, and the mass-radius relation of neutron stars \cite{Men04,Niu10} as well as the cooling mechanism of protoneutron stars \cite{Hua15}. We have no physical reason to preclude possible $\delta$ meson effects in determining the critical density for \D formation in neutron stars. However, to consider effects of the $\delta$ meson one has to introduce all the necessary but unknown $\delta$-baryon coupling constants. Given the fact that our main findings in this work already depend on the completely unknown $\Delta$-$\rho$ coupling, the inclusion of the $\delta$ into our current model is thus not beneficial for the main purposes of this work. 

\section{Acknowledgement}
We would like to thank Lie-Wen Chen, Xiao-Tao He and Xiao-Hua Li for useful discussions.
This work is supported in part by the National Aeronautics and Space Administration
under Grant No. NNX11AC41G issued through the Science Mission
Directorate, the National Science Foundation under Grant No.
PHY-1068022 and PHY-1205019 and the Department of Energy's Office of Science under Award Number DE-SC0013702.

\appendix
\begin{widetext}
\section{Derivation of equation (\ref{def1})}

We start from the relation
$\mu_{\Delta^{-}}=2\mu_{\textrm{n}}-\mu_{\textrm{p}}$, where
 \begin{align}
\mu_{\textrm{n}}=&\sqrt{\left(k^{\textrm{n}}_{\textrm{F}}\right)^2+m_{\textrm{dirac}}^{\ast,2}}
+g_{\omega\textrm{N}}\overline{\omega}_0
-g_{\rho\textrm{N}}\overline{\rho}_0^{(3)},\\
\mu_{\textrm{p}}=&\sqrt{\left(k^{\textrm{p}}_{\textrm{F}}\right)^2+m_{\textrm{dirac}}^{\ast,2}}
+g_{\omega\textrm{N}}\overline{\omega}_0
+g_{\rho\textrm{N}}\overline{\rho}_0^{(3)},\\
\mu_{\Delta^-}=&\sqrt{\left(k^{\Delta^-}_{\textrm{F}}\right)^2+\left(m_{\Delta}-g_{\sigma\Delta}\overline{\sigma}\right)^2}
+g_{\omega\Delta}\overline{\omega}_0
-3g_{\rho\Delta}\overline{\rho}_0^{(3)}
\end{align}
are the chemical potential of neutron, proton and $\Delta^-$, and
$m_{\textrm{dirac}}^{\ast}=m_{\textrm{N}}-g_{\sigma\textrm{N}}\overline{\sigma}$.
The condition to determine $\rho_{\Delta^-}^{\textrm{crit}}$
suggests that we can set $ \rho_{\Delta^-}=0$,
$k_{\textrm{F}}^{\Delta^-}=0$. Then using the relation above we find
that
\begin{align}
&m_{\Delta}-g_{\sigma\Delta}\overline{\sigma}
+g_{\omega\Delta}\omega_0 -3g_{\rho\Delta}\overline{\rho}_0^{(3)}
=\sqrt{\left(k^{\textrm{n}}_{\textrm{F}}\right)^2+m_{\textrm{dirac}}^{\ast,2}}
+g_{\omega\textrm{N}}\omega_0
-g_{\rho\textrm{N}}\overline{\rho}_0^{(3)}+\mu_{\textrm{n}}-\mu_{\textrm{p}},
\end{align}
where the last two terms on the right hand side can be expressed
as\,\cite{XuJ09,Cai12}
\begin{equation}\mu_{\textrm{n}}-\mu_{\textrm{p}}\simeq4E_{\textrm{sym}}(\rho)\delta.
\end{equation}
The expression for the symmetry energy in the nonlinear RMF model
can be written as\,\cite{LCK}
\begin{equation}\label{NEsym}
E_{\textrm{sym}}(\rho)=E_{\textrm{sym}}^{\textrm{kin}}(\rho)+E_{\textrm{sym}}^{\textrm{pot}}(\rho)
 = \frac{k_{\textrm{F}}}{6\sqrt{k_{\textrm{F}}^2+m_{\textrm{dirac}}^{\ast,2}}} + \frac{1}{2}\frac{g_{\rho\textrm{N}}^2\rho}{m_{\rho}^2+\Lambda_{\textrm{V}}g_{\rho\textrm{N}}^2
g_{\omega\textrm{N}}^2\overline{\omega}_0^2}.
\end{equation}
And the equation of motion for $\overline{\rho}_0^{(3)}$-field is
given as
\begin{equation}
m_{\rho}^2\overline{\rho}_0^{(3)}=g_{\rho\textrm{N}}
\left(-\rho\delta-\Lambda_{\textrm{V}}g_{\rho\textrm{N}}\overline{\rho}_0^{(3)}g_{\omega\textrm{N}}^2\overline{\omega}_0^2\right).
\end{equation}
Rearranging this last equation we find
\begin{equation}\label{Nw0}
\overline{\rho}_0^{(3)}=-\frac{g_{\rho\textrm{N}}\rho\delta}{m_{\rho}^2+\Lambda_{\textrm{V}}g_{\rho\textrm{N}}g_{\omega\textrm{N}}^2
\overline{\omega}_0^2}.
\end{equation}
This in turn can be rearranged to give
\begin{align}
-2g_{\rho\textrm{N}}\overline{\rho}_0^{(3)}
=&\frac{2g_{\rho\textrm{N}}^2\rho\delta}{m_{\rho}^2+\Lambda_{\textrm{V}}g_{\rho\textrm{N}}g_{\omega\textrm{N}}^2\overline{\omega}_0^2}
\simeq4E_{\textrm{sym}}^{\textrm{pot}}(\rho)\delta
=4(E_{\textrm{sym}}(\rho)-E_{\textrm{sym}}^{\textrm{kin}}(\rho))\delta
.\end{align} The $\overline{\omega}_0$ in the expression of symmetry
energy (\ref{NEsym}) is evaluated at $\delta=0$, whereas the one
appearing in (\ref{Nw0}) has the value of $\delta$ generally being
nonzero, hence we used ``$\simeq$" symbol in the second line.
Expanding now
$[\left(k^{\textrm{n}}_{\textrm{F}}\right)^2+m_{\textrm{dirac}}^{\ast,2}]^{1/2}$
as
\begin{equation}
\sqrt{\left(k^{\textrm{n}}_{\textrm{F}}\right)^2+m_{\textrm{dirac}}^{\ast,2}}\simeq
\frac{k_{\rm{F}}^{\rm{n},2}}{2m_{\rm{dirac}}^{\ast}}+m_{\rm{dirac}}^{\ast},
\end{equation}
and using the relation above we find
\begin{align}
\frac{k_{\rm{F}}^{\rm{n},2}}{2m_{\rm{dirac}}^{\ast}} =&\Phi_{\Delta}
+g_{\sigma\rm{N}}(1-x_{\sigma})\overline{\sigma}
-g_{\omega\rm{N}}(1-x_{\omega})\overline{\omega}_0
+g_{\rho\rm{N}}(1-x_{\rho})\overline{\rho}_0^{(3)}-4x_{\rho}E_{\rm{sym}}^{\rm{kin}}(\rho)\delta
-4(1-x_{\rho})E_{\rm{sym}}(\rho)\delta\\
\simeq&\Phi_{\Delta}
+g_{\sigma\rm{N}}(1-x_{\sigma})\overline{\sigma}
-g_{\omega\rm{N}}(1-x_{\omega})\overline{\omega}_0
+g_{\rho\rm{N}}\overline{\rho}_0^{(3)}
+6x_{\rho}E_{\rm{sym}}^{\rm{pot}}(\rho)\delta-4E_{\rm{sym}}(\rho)\delta\\
\simeq&\Phi_{\Delta}
+g_{\sigma\rm{N}}(1-x_{\sigma})\overline{\sigma}
-g_{\omega\rm{N}}(1-x_{\omega})\overline{\omega}_0
+(6x_{\rho}-2)E_{\rm{sym}}^{\rm{pot}}(\rho)\delta-4E_{\rm{sym}}(\rho)\delta\\
=&\Phi_{\Delta} +g_{\sigma\rm{N}}(1-x_{\sigma})\overline{\sigma}
-g_{\omega\rm{N}}(1-x_{\omega})\overline{\omega}_0
+(6x_{\rho}-6)E_{\rm{sym}}(\rho)\delta-(6x_{\rho}-2)E_{\rm{sym}}^{\rm{kin}}(\rho)\delta\\
=&\Phi_{\Delta} +g_{\sigma\rm{N}}(1-x_{\sigma})\overline{\sigma}
-g_{\omega\rm{N}}(1-x_{\omega})\overline{\omega}_0
+(6x_{\rho}-6)E_{\rm{sym}}^{\rm{pot}}(\rho)\delta+4E_{\rm{sym}}^{\rm{kin}}(\rho)\delta,
\end{align} where $x_{\phi}=g_{\phi\rm{N}}/g_{\phi\Delta}$ with
$\phi=\sigma$, $\omega$ and $\rho$, and
$\Phi_{\Delta}=m_{\Delta}-m_{\rm{N}}$. This last equation is the Eq.
(\ref{def1}) reported in the main part of this paper. The derivation
of equations (\ref{def2}), (\ref{def3}) and (\ref{def4}) follows
using similar arguments and steps.

\section{Derivation of equation (\ref{eq46})}

Once again we start from the relation
$\mu_{\Delta^{-}}=2\mu_{\textrm{n}}-\mu_{\textrm{p}}$ , and have
\begin{align}
\mu_{\Delta^{-}}=&2\mu_{\textrm{n}}-\mu_{\textrm{p}}=\mu_{\textrm{n}}+\mu_{\textrm{n}}-\mu_{\textrm{p}}
\simeq\mu_{\textrm{n}}+4E_{\textrm{sym}}(\rho)\delta+\mathcal{O}(\delta^3)\end{align}
where $E_{\textrm{sym}}(\rho)$ is the nuclear symmetry energy,
$\delta\equiv1-2x_{\textrm{p}}$ is isospin assymetry, and
$x_{\textrm{p}}$ is the proton fraction. The chemical potential for
neutrons can be expressed as
\begin{equation}
\mu_{\textrm{n}}=\frac{\partial\varepsilon_{\textrm{N}}}{\partial\rho_{\textrm{n}}}=\frac{\partial\varepsilon_{\textrm{N}}}{\partial\rho}
+\frac{2\tau_3^{\textrm{n}}\rho_{\textrm{p}}}{\rho^2}\frac{\partial\varepsilon_{\textrm{N}}}{\partial\delta},\end{equation}
where $\rho$ and $\delta$ are two independent variables. In
obtaining the expression above we have used the relation
\begin{equation}
\frac{\partial}{\partial \rho_J}=\frac{\partial \rho}{\partial
\rho_J}\frac{\partial}{\partial \rho}
+\frac{\partial\delta}{\partial
\rho_J}\frac{\partial}{\partial\delta}=\frac{\partial}{\partial
\rho}+\frac{2\tau_3^J\rho_{\overline{J}}}{\rho^2}\frac{\partial}{\partial\delta}
,
\end{equation}
where $J$=$\{$n, p$\}$,
$\rho_{\overline{\textrm{n}}}=\rho_{\textrm{p}}$,
$\rho_{\overline{\textrm{p}}}=\rho_{\textrm{n}}$. When the density
is smaller than $\rho_{\Delta^-}^{\textrm{cric}}$, there are only
neutrons, protons, electrons and muons present in the system.
Neglecting the contribution from electrons and muons,
$\varepsilon_{\textrm{N}}$ can be taken as the energy density of
nuclear matter only. That is,
\begin{equation}
\varepsilon_{\textrm{N}}(\rho,\delta)=[E(\rho,\delta)+m_{\textrm{N}}]\rho
\,
\end{equation}
where $m_{\textrm{N}}=939\,\textrm{MeV}$ is the static mass of the
nucleon and $E(\rho,\delta)$ is the equation of state of asymmetric
nuclear matter, which in turn can be written as
\begin{equation}
E(\rho,\delta)\simeq
E_0(\rho)+E_{\textrm{sym}}(\rho)\delta^2+\mathcal{O}(\delta^4).\end{equation}
Then,
\begin{align}
\mu_{\Delta^-}\simeq&\mu_{\textrm{n}}+4E_{\textrm{sym}}(\rho)\delta\notag\\
=&E(\rho,\delta)+m_{\textrm{N}}+\rho\frac{\partial
E(\rho,\delta)}{\partial\rho}
+\frac{2\tau_3^{\textrm{n}}\rho_{\textrm{p}}}{\rho^2}
\frac{\partial}{\partial\delta}\left(\rho E_{\textrm{sym}}(\rho)\delta^2\right)+4E_{\textrm{sym}}(\rho)\delta\notag\\
\simeq&E_0(\rho)+E_{\textrm{sym}}(\rho)\delta^2+m_{\textrm{N}}+\rho\frac{\partial
E(\rho,\delta)}{\partial\rho}+\frac{2\tau_3^{\textrm{n}}\rho_{\textrm{p}}}{\rho}
\frac{\partial}{\partial\delta}\left(E_{\textrm{sym}}(\rho)\delta^2\right)+4E_{\textrm{sym}}(\rho)\delta\notag\\
=&E_0(\rho)+E_{\textrm{sym}}(\rho)\delta^2+m_{\textrm{N}}+\rho\frac{\partial
E(\rho,\delta)}{\partial\rho}+4E_{\textrm{sym}}(\rho)\delta(1-x_{\textrm{p}})\notag\\
=&E_0(\rho)+E_{\textrm{sym}}(\rho)\delta^2+m_{\textrm{N}}+\rho\frac{\partial
E(\rho,\delta)}{\partial\rho}
+E_{\textrm{sym}}(\rho)(2\delta+2\delta^2)\notag\\
=&E_0(\rho)+m_{\textrm{N}}+\rho\left[\frac{\partial
E_0(\rho)}{\partial\rho}+\frac{\partial}{\partial\rho}\left(E_{\textrm{sym}}(\rho)\delta^2\right)\right]
+E_{\textrm{sym}}(\rho)(2\delta+3\delta^2)\notag\\
\simeq&E_0(\rho_0)+m_{\textrm{N}}+\left(\frac{1}{2}\chi^2+\frac{\rho}{3\rho_0}\chi\right)K_0
+\left(\frac{1}{6}\chi^3+\frac{\rho}{6\rho_0}\chi^2\right)J_0\notag\\
&+\left(2\delta+3\delta^2\right)E_{\textrm{sym}}(\rho_0)+\left(\left(2\delta+3\delta^2\right)\chi+\frac{\rho}{3\rho_0}\delta^2\right)L(\rho_0),
\end{align}
which is the desired Eq. (\ref{eq46}). Note that $K_0\equiv
K_0(\rho_0)$ and $J_0\equiv J_0(\rho_0)$, and also in the derivation
above we have made use of the following relation,
\begin{equation}
\frac{\partial\chi}{\partial\rho}=\frac{\partial}{\partial\rho}\frac{\rho-\rho_0}{3\rho_0}=\frac{1}{3\rho_0}.
\end{equation}

\end{widetext}



\begin{thebibliography}{99}


\bibitem{Mig78} A.B. Migdal, Rev. Mod. Phys. \textbf{50}, 107 (1978).

\bibitem{Mig90} A.B. Migdal, E.E. Saperstein, M.A. Troitsky, and
D.N. Voskresensky, Phys. Rep. \textbf{192}, 179 (1990).

\bibitem{Erc88} T. Ericson and W. Weise, Pions and Nuclei, Oxford Univ. Press, 1988, ISBN 0-19-852008-5.

\bibitem{VJ} V.R. Pandharipande, Nucl. Phys. \textbf{A178}, 123 (1971).

\bibitem{Bog82} J. Boguta, Phys. Lett. \textbf{B109}, 251 (1982).

\bibitem{Liz97} Z.X. Li, G.J. Mao, Y.Z. Zhuo, and W. Greiner, Phys. Rev. C
\textbf{56}, 1570 (1997).

\bibitem{Oli00} J.C.T. de Oliveira, M. Kyotoku, M. Chiapparini, H.
Rodrigues, and S.B. Duarte, Mod. Phys. Lett. A \textbf{15}, 1529
(2000).

\bibitem{Kos97} D.S. Kosov, C. Fuchs, B.V. Martemyanov, and A. Faessler,
Phys. Lett. \textbf{B421}, 37 (1997).

\bibitem{Sha83} S.L. Shapiro and S.A. Teukolsky, \textit{Black Holes, White Dwarfs, and Neutron
Stars}, Wiley-VCH, 1983, Section 8.11.

\bibitem{Brown75} G.E. Brown and W. Weise, Phys. Rep. \textbf{22}, 279 (1975).

\bibitem{Oset81} E. Oset and A. Plalanques-Mestre, Nucl. Phys. \textbf{A359}, 289 (1981).

\bibitem{Oset87} E. Oset and L.L. Salcedo, Nucl. Phys. \textbf{A468}, 631 (1987).

\bibitem{Con90} J. O'Connell, R. Sealock, Phys. Rev. C \textbf{42}, 2290 (1990).

\bibitem{Jon92} F. de Jong, R. Malfliet, Phys. Rev. C \textbf{46}, 2567 (1992).

\bibitem{Baldo94} M. Baldo, L. Ferreira, Nucl. Phys. \textbf{A569}, 645 (1994).

\bibitem{Gle85} N.K. Glendenning, Astrophys. J. \textbf{293}, 470 (1985).

\bibitem{Gle91} N.K. Glendenning and S. Moszkowski, Phys. Rev. Lett. \textbf{67}, 2414 (1991).

\bibitem{Gle00} N.K. Glendenning, \textit{Compact Stars}, 2nd edition, Spinger-Verlag New York, Inc., 2000.

\bibitem{Xia03} H. Xiang and H. Guo, Phys. Rev. C \textbf{67}, 038801 (2003).

\bibitem{ChY07} Y.J. Chen, H. Guo, and Y. Liu, Phys. Rev. C \textbf{75}, 035806 (2007); Y.J. Chen and H. Guo, Comm. Theor. Phys.
\textbf{49}, 1283 (2008); Y.J. Chen, Y. Yuan, and Y. Liu, Phys. Rev.
C \textbf{79}, 055802 (2009).

\bibitem{Sch10} T. Schurhoff, S. Schramm, and V. Dexheimer, Astrophys. J. \textbf{724}, L74 (2010).

\bibitem{Lav10} A. Lavagno, Phys. Rev. C \textbf{81}, 044909 (2010).

\bibitem{Dra14} A. Drago, A. Lavagno, and G. Pagliara, Phys. Rev. D \textbf{89}, 043104 (2014).

\bibitem{Dra14a} A. Drago, A. Lavagno, G. Pagliara, and D. Pigato, (2014) [arXiv:1407.2843] and references therein.

\bibitem{Esk98} M. Eskef et al. (FOPI collaboration)  Eur. Phys. J. A3 (1998) 335 (1998).

\bibitem{Kor04} C. L. Korpa and A. E. L. Dieperink, Phys. Rev. C{\bf 70}, 015207 (2004).

\bibitem{Hee05} Hendrik van Hees and Ralf Rapp, Phys. Lett. B{\bf 606}, 59 (2005).

\bibitem{EPJA} B.A. Li, A. Ramos, G. Verde, and I. Vida\~na, eds., ``Topical issue on nuclear symmetry energy", Eur. Phys. J. A \textbf{50}, No. 2, (2014).

\bibitem{lkb98} B.A. Li, C.M. Ko, and W. Bauer, topical review, Int. Jou. Mod. Phys. E \textbf{7}, 147 (1998).

\bibitem{liudo}{\it Isospin Physics in Heavy-Ion Collisions at Intermediate Energies}, Eds. Bao-An Li and W. Udo Schr\"oder, ISBN 1-56072-888-4,
Nova Science Publishers, Inc (2001, New York).

\bibitem{Bar05} V. Baran, M. Colonna, V. Greco, and M.Di Toro, Phys. Rep.
\textbf{410}, 335 (2005).

\bibitem{LCK} B.A. Li, L.W. Chen, and C.M. Ko, Phys. Rep. \textbf{464}, 113 (2008).

\bibitem{Trau12} W. Trautmann and H.H. Wolter, Int. J. Mod. Phys. E  \textbf{21}, 1230003 (2012).

\bibitem{Tsang12} M.B. Tsang, \textit{et al.}, Phys. Rev. C  \textbf{86}, 015803 (2012).

\bibitem{LiBA13} B.A. Li, X. Han, Phys. Lett. \textbf{B727}, 276 (2013).

\bibitem{Hor14} C.J. Horowitz \textit{et al.}, J. of Phys. G  \textbf{41}, 093001 (2014).

\bibitem{Jim13} J.M. Lattimer, Annu. Rev. Nucl. Part. Sci. \textbf{62}, 485 (2012).

\bibitem{HIon} H. St$\ddot{\textbf{o}}$cker and W. Greiner, Phys. Rep.
\textbf{137}, 277 (1986); G.F. Bertsch and S.D. Gupta, Phys. Rep.
\textbf{160}, 189 (1988); W. Cassing, V. Metag, U. Mosel, K. Niita
Physics Reports \textbf{188}, 363 (1990); J. Aichelin, Phys. Rep.
\textbf{202}, 233 (1991); S.A. Bass, \textit{et al.}, Prog. Nucl.
Part. Phys. \textbf{41}, 255 (1998); W. Cassing and E.L.
Bratkovskaya, Phys. Rep. \textbf{308}, 65 (1999); B.A. Li, C.M. Ko,
A.T. Sustich and B. Zhang, Int. Journal of Modern Physics
\textbf{10}, 267 (2001); O. Buss, \textit{et al.}, Phys. Rep.
\textbf{512}, 1 (2012).

\bibitem{LiLisa} B.A. Li and M. Lisa, manuscript (2014).

\bibitem{Lenske} H. Lenske, talk at the NUSTAR Annual Collaboration Meeting, March 3-7, 2014, \\
\url{indico.gsi.de/conferenceOtherViews.py?view=standard&confId=2354}

\bibitem{Ser86} B.D. Serot and J.D. Walecka, Adv. Nucl. Phys. \textbf{16}, 1 (1986); Int. J. Mod. Phys. E \textbf{6}, 515 (1997).

\bibitem{Rei89} P.G. Reinhard, Rep. Prog. Phys. \textbf{52}, 439 (1989).

\bibitem{Rin96} P. Ring, Prog. Part. Nucl. Phys. \textbf{37}, 193 (1996).

\bibitem{Men06} J. Meng, H. Toki, S.G. Zhou, S.Q. Zhang, W.H. Long, and L.S.
Geng, Prog. Part. Nucl. Phys. \textbf{57}, 470 (2006).

\bibitem{Sug94} Y. Sugahara and H. Toki, Nucl. Phys. \textbf{A579}, 557 (1994).

\bibitem{Lal97} G.A. Lalazissis, J. K$\ddot{\textbf{o}}$nig, P. Ring, Phys. Rev. C \textbf{55}, 540 (1997).

\bibitem{Lon04} W.H. Long, J. Meng, N.Van Giai, and S.G. Zhou, Phys. Rev. C \textbf{69}, 034319 (2004).

\bibitem{Jia10} W.Z. Jiang, Phys. Rev. C \textbf{81}, 044306 (2010).

\bibitem{Fat10} F.J. Fattoyev, C.J. Horowitz, J. Piekarewicz, and G. Shen, Phys. Rev. C
\textbf{82}, 055803 (2010).

\bibitem{Fat10a} F.J. Fattoyev and J. Piekarewicz Phys. Rev. C
\textbf{82}, 025805 (2010); \textbf{82}, 025810 (2010); \textbf{84},
064302 (2011); \textbf{86}, 105802 (2012).

\bibitem{Agr12} B.K. Agrawal, A. Sulaksono, and P.-G. Reinhard, Nucl. Phys. \textbf{A882}, 1 (2012).

\bibitem{Fat13} F.J. Fattoyev, J. Carvajal, W.G. Newton, and B.A. Li, Phys. Rev. C
\textbf{87}, 015806 (2013).

\bibitem{Mul96} H. M\"{u}ller and B.D. Serot, Nucl. Phys.
\textbf{A606}, 508 (1996).

\bibitem{Hor01} C.J. Horowitz, and J. Piekarewicz, Phys. Rev. Lett \textbf{%
86}, 5647 (2001); Phys. Rev. C \textbf{64}, 062802(R) (2001); Phys.
Rev. C \textbf{66}, 055803 (2002).

\bibitem{Tod05} B.G. Todd-Rutel and J. Piekarewicz, Phys. Rev. Lett.
\textbf{95}, 122501 (2005).

\bibitem{Che07} L.W. Chen, C.M. Ko, and B.A. Li, Phys. Rev. C \textbf{76}, 054316 (2007).

\bibitem{Cai12} B.J. Cai and L.W. Chen, Phys. Rev. C \textbf{85},
024302 (2012).

\bibitem{Cai14} B.J. Cai and L.W. Chen, (2014) [arXiv:1402.4242].

\bibitem{Cai14a} B.J. Cai, L.W. Chen, and W. Z. Jiang, in
preparation (2014).

\bibitem{Fra81} L.L. Frankfurt and M.I. Strikman, Phys. Rep. \textbf{76}, 215 (1981); Phys. Rep. \textbf{160}, 235 (1988);
L.L. Frankfurt, M. Sargsian, and M.I. Strikman, Int. Mod. Phys. A
\textbf{23}, 2991 (2008).

\bibitem{Arr12} J. Arrington, D.W. Higinbotham, G. Rosner, and M.
Sargsian, Prog. Part. Nucl. Phys. \textbf{67}, 898 (2012).


\bibitem{Hen14} O. Hen, \textit{et al.}, Science \textbf{346}, 614 (2014).

\bibitem{XuLi} C. Xu and B.A. Li, arXiv: 1104.2075; C. Xu, A. Li, B.A. Li, J. of Phys: Conference Series \textbf{420}, 012190 (2013).

\bibitem{Vid11} I. Vidana, A. Polls, C. Providencia, Phys. Rev. C 84, 062801(R) (2011).

\bibitem{Lov11} A. Lovato, O. Benhar, S. Fantoni, A.Yu. Illarionov, and K.E. Schmidt, Phys. Rev. C \textbf{83}, 054003 (2011).

\bibitem{Car12} A. Carbone, A. Polls, A. Rios, Eur. Phys. Lett. \textbf{97}, 22001 (2012).

\bibitem{Rio14} A. Rios, A. Polls, W.H. Dickhoff, Phys. Rev. C \textbf{89}, 044303 (2014).

\bibitem{OrHen14} O. Hen, B.A. Li, W.J. Guo, L.B. Weinstein, and E. Piasetzky, Phys.
Rev. C \textbf{91}, 025803 (2015).

\bibitem{LiBA14} B.A. Li, W.J. Guo, and Z.Z. Shi, Phys. Rev. C{\bf 91}, 044601 (2015).

\bibitem{Antoniadis2013} J. Antoniadis \textit{et al.},
Science \textbf{340}, 6131 (2013).

\bibitem{Kit} Y. Kitazoe, M. Sano, H. Toki, and S. Nagamiya, Phys. Lett. \textbf{B166}, 35 (1986).

\bibitem{Li93} B.A. Li, Nucl. Phys. \textbf{A552}, 605 (1993).

\bibitem{Hae07} P. Haensel, A.Y. Potekhin, and D.G. Yakovlev, \textit{Neutron Stars 1}, Springer, 2007.

\bibitem{Cai12a} B.J. Cai and L.W. Chen, Phys. Lett.
\textbf{B711}, 104 (2012).

\bibitem{Che14a} W.C. Chen and J. Piekarewicz, Phys. Rev. C
\textbf{90}, 044305 (2014).

\bibitem{Che10} L.W. Chen, C.M. Ko, B.A. Li, and J. Xu, Phys. Rev. C
\textbf{82}, 024321 (2010).

\bibitem{Zha13} Z. Zhang and L.W. Chen, Phys. Lett. \textbf{B726}, 234 (2013).

\bibitem{Che12} L.W. Chen and J.Z. Gu, J. Phys. G \textbf{39}, 035104 (2012).

\bibitem{Che11a} L.W. Chen, Sci. China: Phys. Mech. Astron. \textbf{54},
suppl. 1, s124 (2011) [arXiv:1101.2384].


\bibitem{Zha14} Z. Zhang and L.W. Chen, Phys. Rev. C \textbf{90},
064317 (2014).

\bibitem{Che11} L.W. Chen, Phys. Rev. C \textbf{83}, 044308 (2011).


\bibitem{XuJ09} J. Xu, L.W. Chen, B.A. Li, and H.R. Ma, Phys. Rev. C \textbf{%
79}, 035802 (2009); Astrophys. J. \textbf{697}, 1549 (2009).

\bibitem{Hor03} J. Carriere, C. J. Horowitz, and J. Piekarewicz, Astrophys.
J. \textbf{593}, 463 (2003).

\bibitem{BPS71} G. Baym, C. Pethick, and P. Sutherland, Astrophys. J. \textbf{%
170}, 299 (1971); K. Iida and K. Sato, Astrophys. J. \textbf{477},
294 (1997).

\bibitem{Yak01} D.G. Yakovlev, A.D. Kaminker, O.Y. Gnedin, and P.
Haensel, Phys. Rep. \textbf{354}, 1 (2001); D. Page, J.M. Lattimer,
M. Prakash, and A. Steiner, Astrophys. J. Supp. Ser. \textbf{155},
623 (2004); D. Page, U. Geppert, and F. Weber, Nucl. Phys.
\textbf{A777}, 479 (2006); D. Page and S. Reddy, Annu. Rev. Nucl.
Part. Sci. \textbf{56}, 327 (2006).

\bibitem{Kub97} S. Kubis and M. Kutschera, Phys. Lett. B{\bf 399}, 191 (1997).

\bibitem{Liu02} B. Liu, V. Greco, V. Baran, M. Colonna, and M. Di Toro, Phys. Rev. C {\bf 65}, 045201 (2002).

\bibitem{Gre03} V. Greco, M. Colonna, M. Di Toro, and F. Matera, Phys. Rev. C {\bf 67}, 015203 (2003).

\bibitem{Men04} D. P. Menezes and C. Providência, Phys. Rev. C {\bf 70}, 058801 (2004).

\bibitem{Gai04} T. Gaitanos, M. Di Toro, S. Typel, V. Baran, C. Fuchs, V. Greco, and H. H. Wolter, Nucl. Phys. A {\bf 732}, 24 (2004).

\bibitem{Ala10} A. G. Alaverdyan, G. B. Alaverdyan, A. O. Chiladze,
Int. Journal of Modern Phys. D{\bf 19}, 1557 (2010).

\bibitem{Ala10a} G. B. Alaverdyan, Res. Astron. Astrophys.10:1255 (2010).

\bibitem{Pai09} Helena Pais, Alexandre Santos, Constança Providência, Phys. Rev. C{\bf 80}, 045808 (2009).

\bibitem{Niu10} Z.M. Niu and C.Y.  Gao, Int. J. Mod. Phys. E19, 2247 (2010).


\bibitem{Hua15} X.L. Huang et al., arXiv:1502.01460. 


\end{thebibliography}
\end{document}